\documentclass[apj, numberedappendix]{emulateapj}

\usepackage{amssymb, amsmath}
\usepackage{color}

\shorttitle{Angular momentum profile for haloes}
\shortauthors{Liao, Chen \& Chu}

\begin{document}
\title{A universal angular momentum profile for dark matter haloes}
\author{Shihong Liao, Jianxiong Chen, and M.-C. Chu}
\affil{Department of Physics and Institute of Theoretical Physics, The Chinese University of Hong Kong, Shatin, N.T., Hong Kong; liaoshong@gmail.com, jxchen@phy.cuhk.edu.hk, mcchu@phy.cuhk.edu.hk}

\begin{abstract}
The angular momentum distribution in dark matter haloes and galaxies is a key ingredient in understanding their formation. Especially, the internal distribution of angular momenta is closely related to the formation of disk galaxies. In this article, we use haloes identified from a high-resolution simulation, the Bolshoi simulation, to study the spatial distribution of specific angular momenta, $j(r,\theta)$. We show that by stacking haloes with similar masses to increase the signal-to-noise ratio, the profile can be fitted as a simple function, $j(r,\theta)=j_s \sin^2(\theta/\theta_s) (r/r_s)^2/(1+r/r_s)^4 $, with three free parameters, $j_s, r_s$, and $\theta_s$. Specifically, $j_s$ correlates with the halo mass $M_\mathrm{vir}$ as $j_s\propto M_\mathrm{vir}^{2/3}$, $r_s$ has a weak dependence on the halo mass as $r_s \propto M_\mathrm{vir}^{0.040}$, and $\theta_s$ is independent of $M_\mathrm{vir}$. This profile agrees with that from a rigid shell model, though its origin is unclear. Our universal specific angular momentum profile $j(r,\theta)$ is useful in modelling haloes' angular momenta. Furthermore, by using an empirical stellar mass - halo mass relation, we can infer the averaged angular momentum distribution of a dark matter halo. The specific angular momentum - stellar mass relation within a halo computed from our profile is shown to share a similar shape as that from the observed disk galaxies.
\end{abstract}

\keywords{galaxies: formation - galaxies: halos - galaxies: dwarf - cosmology: dark matter}

\section{Introduction}
The angular momentum distribution of matter in haloes and galaxies plays an important role in their formation \citep[see][ for a recent review]{schafer2009}. Especially, the distribution of angular momenta inside a halo is directly related to the density profile of the galactic disc \citep[see \S 11.4 of][ and references therein]{mo2010}. Dark matter haloes in equilibrium are expected to share some universal properties. For example, it was shown in cosmological N-body simulations that virialized haloes follow a universal density profile $\rho(R)$, $R$ being the radial distance from center, which can be fitted as a simple function with two free parameters $\rho_s$ and $R_s$, $\rho(R)=\rho_s/(R/R_s)(1+R/R_s)^2$ \citep[the NFW profile,][]{navarro1995,navarro1996,navarro1997}. A natural and interesting question is whether haloes' angular momenta follow any universal profile. We will address this question in this article.

\citet[]{barns1987} extract the \textit{differential} specific angular momentum profile from N-body simulations, and show that $j_d(R)\propto R$, where $j_d(R)$ is the specific angular momentum of a spherical shell with radius $R$. \citet[]{bett2010} look at the \textit{cumulative} specific angular momentum profile for a halo and conclude that $j_c(<R)\propto R$, where $j_c(<R)$ is the specific angular momentum inside $R$. \citet[]{bullock2001} present a universal \textit{mass profile} that is related to the angular momenta of galactic haloes, $M_\mathrm{emp}(<j)=M_\mathrm{vir}\mu (j/j_\mathrm{max})/(\mu-1+j/j_\mathrm{max})$, where $M_\mathrm{emp}(<j)$ is the mass with specific angular momenta smaller than $j$, $M_\mathrm{vir}$ is the halo virial mass, $j_\mathrm{max}$ is the maximum specific angular momentum inside the halo, and $\mu$ is a free parameter; see also \citet[]{bosch2002}, \citet[]{chen2002}, \citet[]{chen2003} and \citet[]{sharma2005} for further discussions.

The differential and cumulative specific angular momentum profiles mentioned above were obtained assuming spherical symmetry. However, the angular momentum of a halo defines a special direction, and deviations from spherical symmetry should be taken into account. \citet[]{bullock2001} discuss the cylindrically symmetric spatial profile of angular momenta for individual haloes, and notice that haloes tend to have larger $j$ in the equatorial plane and smaller $j$ along the polar direction. However, the noises in their profiles of individual haloes, which are caused by substructures and complicated formation histories, are fairly large. 

In this article, we reduce such noises by stacking high-resolution haloes with similar masses. The stacking method significantly increases the signal-to-noise ratio and is widely used in astronomical image processing and studying haloes' \citep[e.g.][]{gao2008, hayashi2008, reed2011} and voids' \citep[]{hamaus2014} density profiles from cosmological N-body simulations. 

As pointed out by \citet[]{bullock2001} and \citet[]{chen2002}, the measurements of haloes' angular momenta are affected by the discreteness effects introduced by particle sampling. To probe the spatial profile of $j$, we need simulations with very high resolution. In this article, we use the Bolshoi simulation \citep[]{klypin2011}, which has high mass and force resolutions, to study the $j-$profile of dark matter haloes. 

The article is structured as follows. We describe the halo sample from the Bolshoi simulation and analysis methods in Section \ref{sec_method}. A universal profile of $j(r,\theta)$ is presented in Section \ref{sec_profile}. In Section \ref{sec_model}, the profile is shown to be qualitatively similar to the one derived from the rigid shell model. In Section \ref{sec_obs}, we compare our profile with observational data from disk galaxies. A discussion and summary is presented in Section \ref{sec_dis}.

\section{Methodology} \label{sec_method}
\subsection{Bolshoi Simulation and Halo Samples}
The Bolshoi simulation\footnote{www.cosmosim.org} uses $2048^3$ particles to sample the phase space distribution of dark matter fluids in a periodic cube (box size $L=250$ $h^{-1}\mathrm{Mpc}$) with the WMAP5 cosmology. The mass and force resolutions are $1.35\times 10^8$ $h^{-1}M_\odot$ and $1.0$ $h^{-1}\mathrm{kpc}$ respectively. A detailed description of the simulation can be found in \citet[]{klypin2011}.  

We use distinct haloes with masses $M=[4, 64]\times 10^{12}h^{-1}M_\odot$ in the Bolshoi BDMV halo catalogue, which are identified using the Bound Density Maximum method \citep[BDM,][]{klypin1997} with the overdensity criterion of $360\rho_\mathrm{back}$. Here $\rho_\mathrm{back}$ is the background matter density. This definition is equivalent to an overdensity parameter $\Delta_\mathrm{vir}=97.2$ with respect to the critical density $\rho_\mathrm{cri}$ \citep[][]{bryan1998}. Note that the Bolshoi database does not provide lists of particles for haloes, and we have to query particles directly from simulation snapshots. From each halo centre, we select all particles inside the halo virial radius $R_\mathrm{vir}$ and regard them as primary halo particles. Then the unbound particles in the primary set are removed using an iteration method (see Appendix \ref{appendix_unbound} for details). The remaining particles are processed to study the angular momentum profile. We only consider haloes at $z=0$ (snapshot 416 in the Bolshoi simulation) in this article.

There are totally 13606 haloes in our sample. The smallest halo in the sample contains $\sim 30000$ particles, which has high enough resolution for us to probe the angular momentum profile. The haloes are divided into 6 mass bins for stacking. The details are summarised in the first three columns of Table \ref{bolshoi_amp}.

\subsection{Calculating and Fitting Methods}
We define the direction of the specific angular momentum of a halo, $\pmb{j}_\mathrm{halo}$, as the $z-$axis and assume that the spatial distribution of $j$ is cylindrically symmetric around it, where $j$ is the $z-$component of the specific angular momentum $\pmb{j}$ at different locations. Under such an assumption, the specific angular momentum profile does not depend on the azimuthal angle $\phi$, and it is a function of $r$ and $\theta$ only, $\theta$ being the zenith angle. We further assume that $j(r,\theta)$ is symmetric between the northern and southern hemispheres, and thus the range of $\theta$ we study reduces to $[0,\pi/2]$.

To calculate $j(r,\theta)$ for a halo, we divide $r\equiv R/R_\mathrm{vir}$ and $\theta$ into several bins requiring that in each bin, the number of particles reaches a threshold of $N_\mathrm{th}$. We adopt $N_\mathrm{th} = 200$ in the following; see Appendix \ref{appendix_fitting} for tests with other thresholds. Note that $r$ is rescaled with $R_\mathrm{vir}$, and its range is from 0 to 1.

Within the $i-$th bin, the specific angular momentum $j_i(r,\theta)$ is computed as
\begin{equation}
j_i(r, \theta)=\frac{\sum_{k=1}^{N_i} m_k \pmb{R}_k \times \pmb{v}_k}{\sum_{k=1}^{N_i}m_k} \cdot \pmb{\hat{z}},
\end{equation}
where $\pmb{R}_k$ and $\pmb{v}_k$ are the position and velocity of particle $k$ with respect to the halo's centre-of-mass position $\pmb{R}_c$ and velocity $\pmb{v}_c$, $m_k$ is the mass of particle $k$, and $N_i$ is the number of particles inside the $i-$th bin. The norm $\pmb{\hat{z}}$ indicates that we use the $z-$axis projection here and consider only the angular momentum of each spatial bin along the direction of the total angular momentum.

For haloes in the same mass bin, we use the same spatial bin dividing scheme for $r$ and $\theta$. Then we stack $j_i(r, \theta)$ from different haloes in the same mass bin to obtain an average profile,
\begin{equation}
\bar{j}_i(r,\theta)=\frac{\sum_{k=1}^{N_h} j_i(r,\theta)}{N_h},
\end{equation}
where $N_h$ is the number of haloes in the mass bin. 

To estimate the errors of $\bar{j}_i(r,\theta)$ in the $i$-th spatial bin, $\sigma_{\bar{j},i}$, we use
\begin{equation}
\sigma_{\bar{j},i}^2 = \sigma_{\mathrm{stat},i}^2 + \sigma_{\mathrm{syst},i}^2,
\end{equation}
where $\sigma_{\mathrm{stat},i}$ and $\sigma_{\mathrm{syst},i}$ are the statistical and systematic errors respectively.
The statistical error accounts for the dispersion of the stacking sample, and is computed as $\sigma_{\mathrm{stat},i}=\sigma_{j,i}/\sqrt{N_h}$, where $\sigma_{j,i}$ is the standard deviation of $j_i(r,\theta)$. The systematic error, $\sigma_{\mathrm{syst},i}$, originates from the numerical discreteness of particles in each spatial bin. To estimate $\sigma_{\mathrm{syst},i}$, we pick the most massive halo from our catalogue, and randomly select a fraction of particles from it to create a low-resolution counterpart. By generating $N_{h,\mathrm{low}}$ such low-resolution realizations, we can compute the mean $\bar{j}_{i,\mathrm{low}}$ and standard deviation $\sigma_{\bar{j},\mathrm{low}}$ in each spatial bin. We find that
\begin{equation}
\sigma_{\bar{j}, \mathrm{low}}\approx \sqrt{\bar{j}_{i,\mathrm{low}}/N_{p,i}}
\end{equation}    
with $N_{p,i}$ being the particle number in the spatial bin. Here, the physical unit of $\bar{j}_{i,\mathrm{low}}$ is $h^{-1}\mathrm{Mpc}$ $\mathrm{km}$ $\mathrm{s}^{-1}$. It implies that a spatial bin with a larger particle number and specific angular momentum tends to estimate $\bar{j}_{i}$ more precisely, i.e., $\sigma_{\bar{j},\mathrm{low}}/\bar{j}_{i,\mathrm{low}} \ll 1$. This is quite similar to the error estimation proposed by \citet[][]{chen2002} based on physical arguments (see their Eq. 4). In this article, we adopt $\sigma_{\mathrm{syst},i} = \sqrt{\bar{j}_{i}/N_{p,i}}$.

The profile is fitted using non-linear least squares by minimizing the residuals with the Levenberg-Marquardt method \citep[]{levenberg1944, marquardt1963}:
\begin{equation}
\chi^2 = \sum\limits_{i=1}^{N_\mathrm{bins}} \frac{\left[\bar{j}_i - j_\mathrm{mod}(r_i, \theta_i)\right]^2}{\sigma_{\bar{j},i}^2},
\end{equation}
where $N_\mathrm{bins}$ is the total number of spatial bins in $r$ and $\theta$, and $j_\mathrm{mod}(r,\theta)$ is the model profile.

\begin{figure}
\includegraphics[width=240pt]{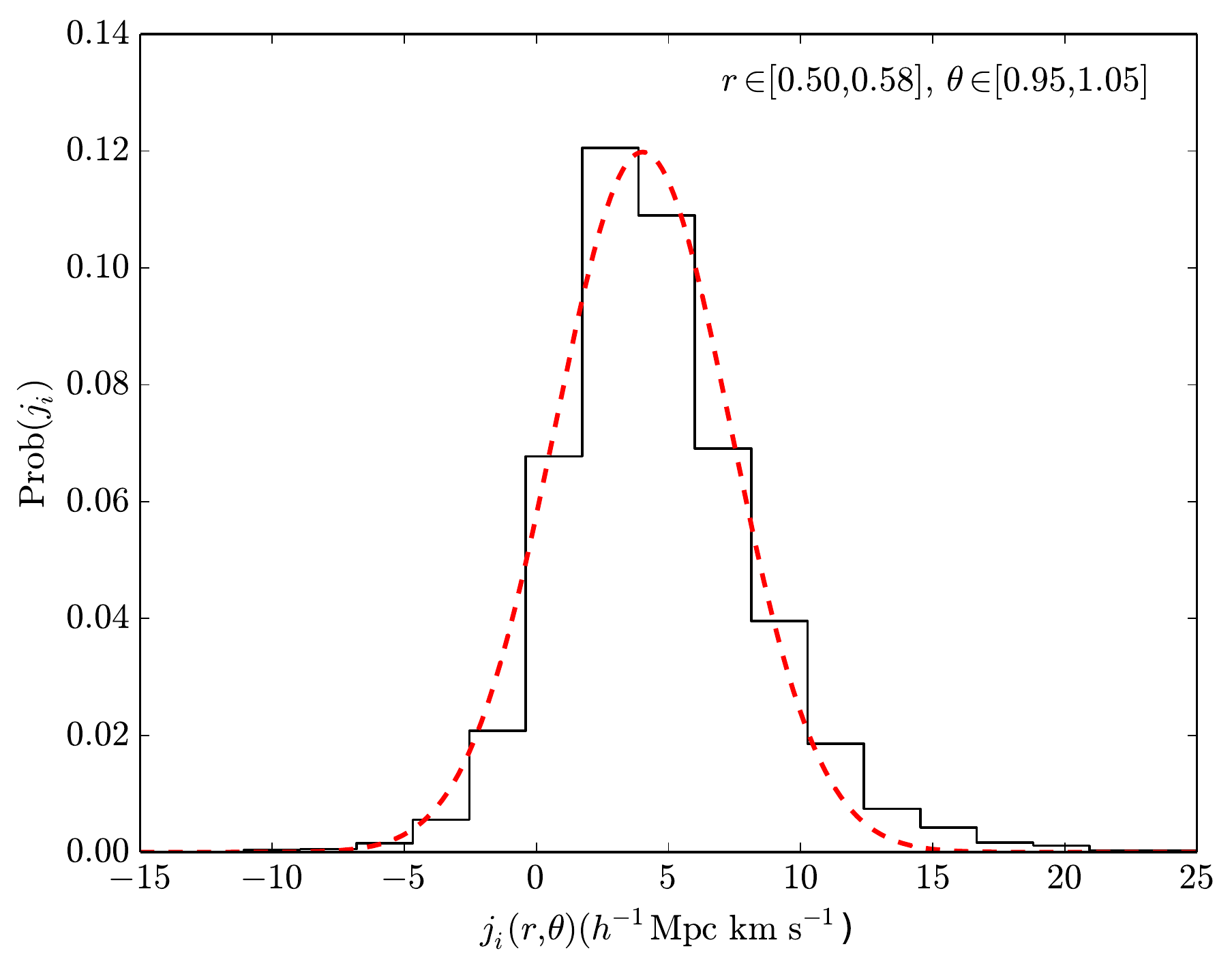}
\caption{Probability distribution of $j_i(r,\theta)$ for the spatial bin of $r\in [0.50, 0.58]$ and $\theta \in [0.95, 1.05]$ in the mass bin of $M_\mathrm{vir} =[4, 4\sqrt{2}]\times 10^{12}$ $h^{-1}M_\odot$. The red dashed line shows a Gaussian distribution function for comparison. Other spatial bins give similar distributions.}
\label{bin_gaussian}
\end{figure}

The fitting method we used assumes the distribution of $j_i(r,\theta)$ in each mass bin to be Gaussian. This can be verified from the data, as shown in Figure \ref{bin_gaussian}. The angular momentum of each spatial bin is a consequence of a series of complicated ``random'' processes (e.g. tidal torquing, collapsing, merging, etc.), and $j_i$ approximate to a normal distribution according to the central limit theorem.

We use the jackknife resampling to estimate the standard errors of the fitted parameters. For a mass bin with $N_h$ haloes, we create $N_h$ resamples by leaving out one halo each time. For each resample with $(N_h-1)$ haloes, we  perform a non-linear least square fit outlined above to the stacked profile. Finally, the standard errors are computed from the $N_h$ estimations of the fitted parameters.

\section{Specific angular momentum profile}\label{sec_profile}
\begin{figure*}\centering
\includegraphics[width=425pt]{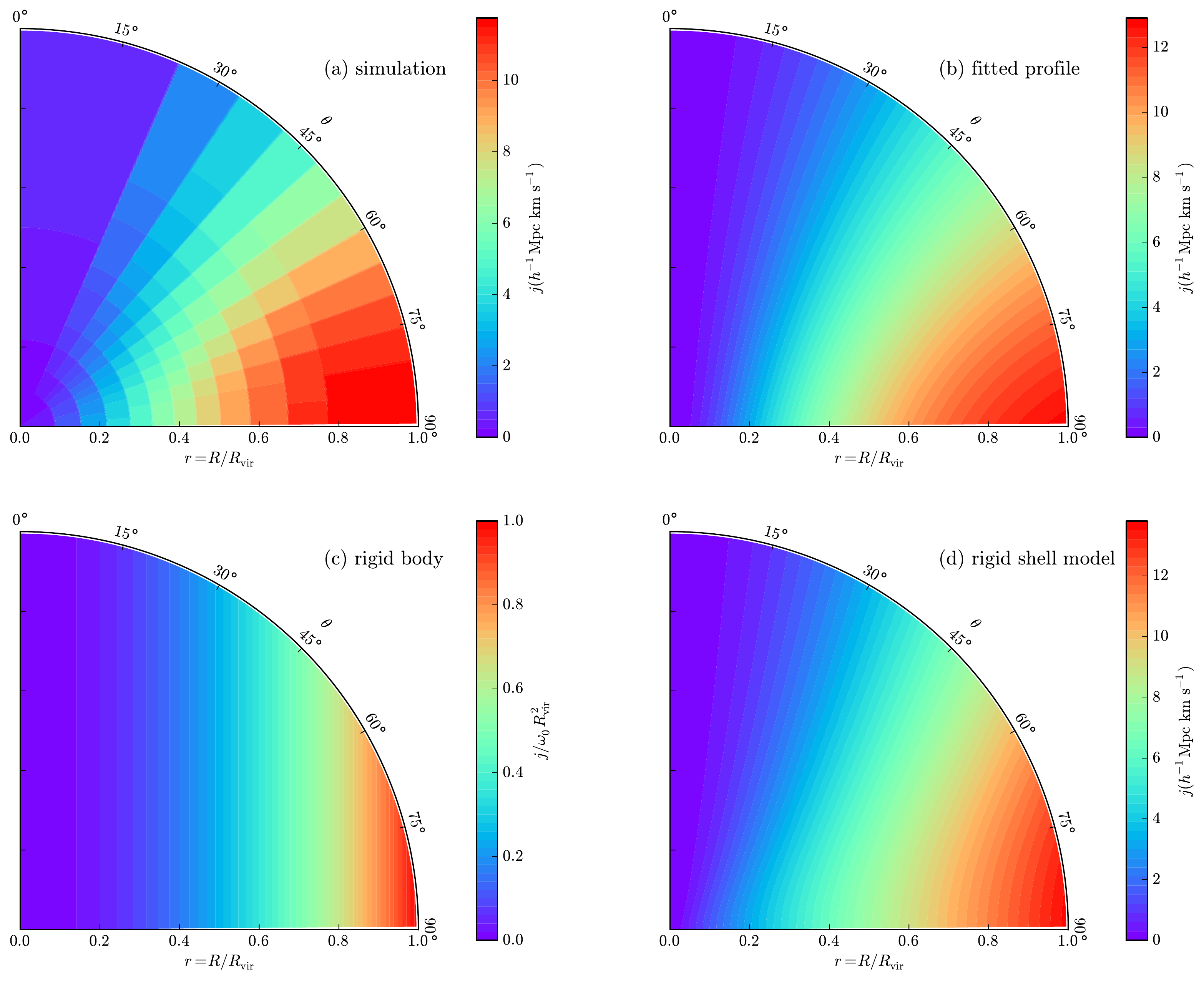}
\caption{Stacked profiles of specific angular momenta: (a) $j(r,\theta)$ computed from the simulated haloes with mass $M_\mathrm{vir}=[4,4\sqrt{2}]\times 10^{12}$ $h^{-1}M_\odot$; (b) the fitted $j_\mathrm{mod}(r,\theta)$ of panel (a) with Equation (\ref{eq_fitted_amp}); (c) $j(r,\theta)$ from a rigid body rotating with a constant angular velocity $\omega_0$; (d) $j(r,\theta)$ from the rigid shell model calculated with an NFW density profile.}
\label{profile_contour}
\end{figure*}

The stacked specific angular momentum profile is shown in panel (a) of Figure \ref{profile_contour}. It can be fitted by an empirical model,
\begin{equation}\label{eq_fitted_amp}
j_\mathrm{mod}(r,\theta)=j_s\frac{(r/r_s)^2}{(1+r/r_s)^4}\sin^2(\theta/\theta_s),
\end{equation}
where $j_s, r_s,$ and $\theta_s$ are free parameters. The best-fits for $j_\mathrm{mod}(r,\theta)$ of different mass bins are summarised in Table \ref{bolshoi_amp}. An example of the fitted profiles is plotted in panel (b) of Figure \ref{profile_contour}.

\begin{figure*}\centering
\includegraphics[width=425pt]{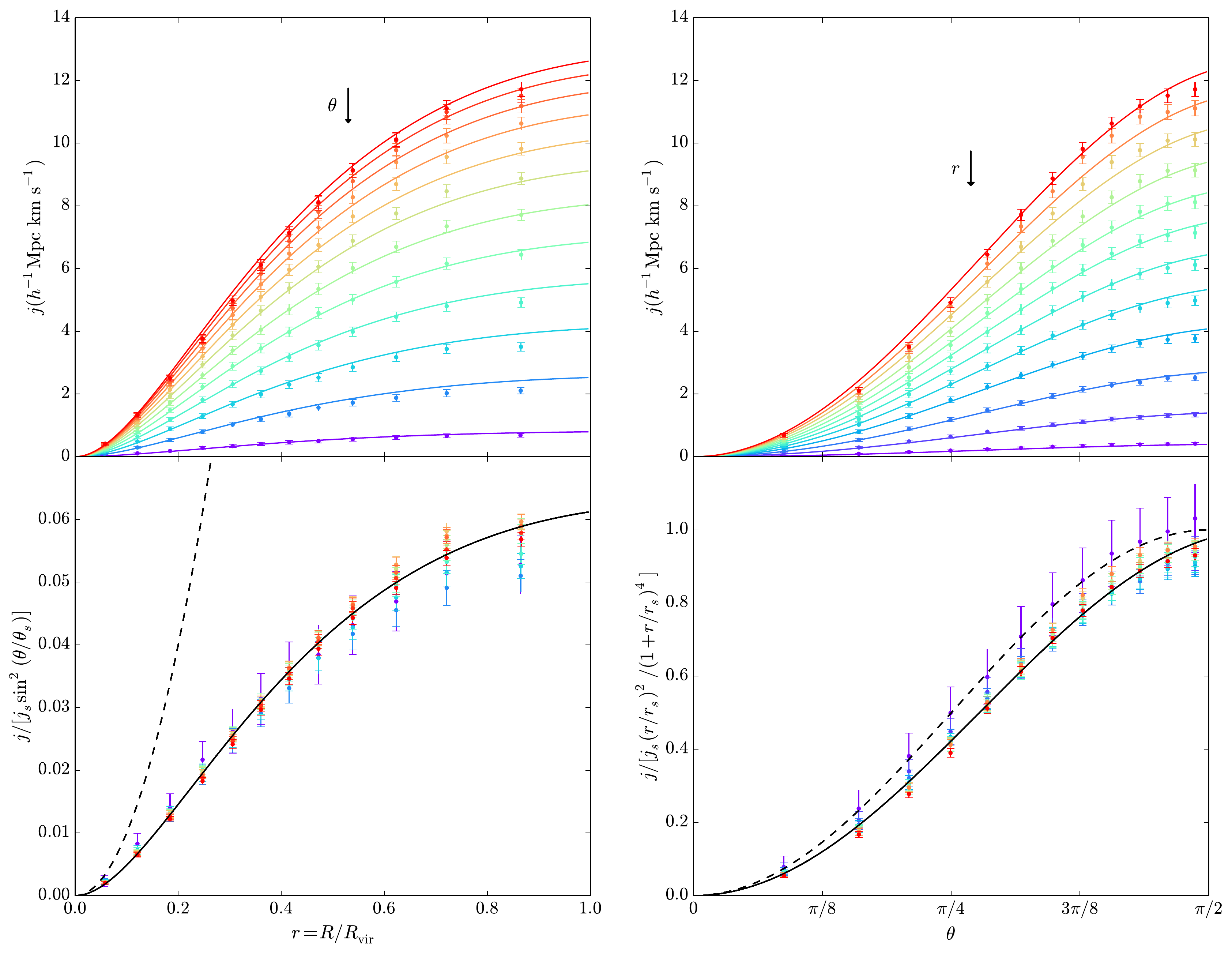}
\caption{Best-fits for the specific angular momentum profiles $j(r,\theta)$ from stacking haloes with $M_\mathrm{vir}=[4, 4\sqrt{2}]\times 10^{12}$ $h^{-1}M_\odot$. The upper left panel shows the best-fits for $j-r$ relations of different $\theta$ bins, and the upper right panel shows $j-\theta$ relations of different $r$ bins. From top to bottom, the value of $\theta$ (or $r$) decreases. In the lower panels, $j$ are rescaled by $j_s \sin^2(\theta/\theta_s)^2$ (left), and $j_s (r/r_s)^2/(1+r/r_s)^4$ (right). The solid lines in the lower panels represent the rescaled relations of $j_\mathrm{res}(r)=(r/r_s)^2/(1+r/r_s)^4$ (left) and $j_\mathrm{res}(\theta) = \sin^2(\theta/\theta_s)$ (right). The dashed lines show the rescaled angular momentum profiles for a rigid body, i.e., $j_\mathrm{RB,res}(r)=r^2$ (left) and $j_\mathrm{RB,res}(\theta)=\sin^2\theta$ (right).}
\label{rescale}
\end{figure*}

The $j-r$ ($j-\theta$) relations for different $\theta$ ($r$) bins of stacked haloes with $M_\mathrm{vir}=[4,4\sqrt{2}]\times 10^{12}$ $h^{-1}M_\odot$ are plotted in the upper left (right) panel of Figure \ref{rescale}. After being rescaled by $j_s \sin^2(\theta/\theta_s)$ $[j_s (r/r_s)^2/(1+r/r_s)^4]$, different $j-r$ ($j-\theta$) relations approximately fall into the same curve; see the lower left (right) panel of Figure \ref{rescale}. This confirms the validity of our proposed fitting model, Equation (\ref{eq_fitted_amp}).

\begin{figure}
\includegraphics[width=250pt]{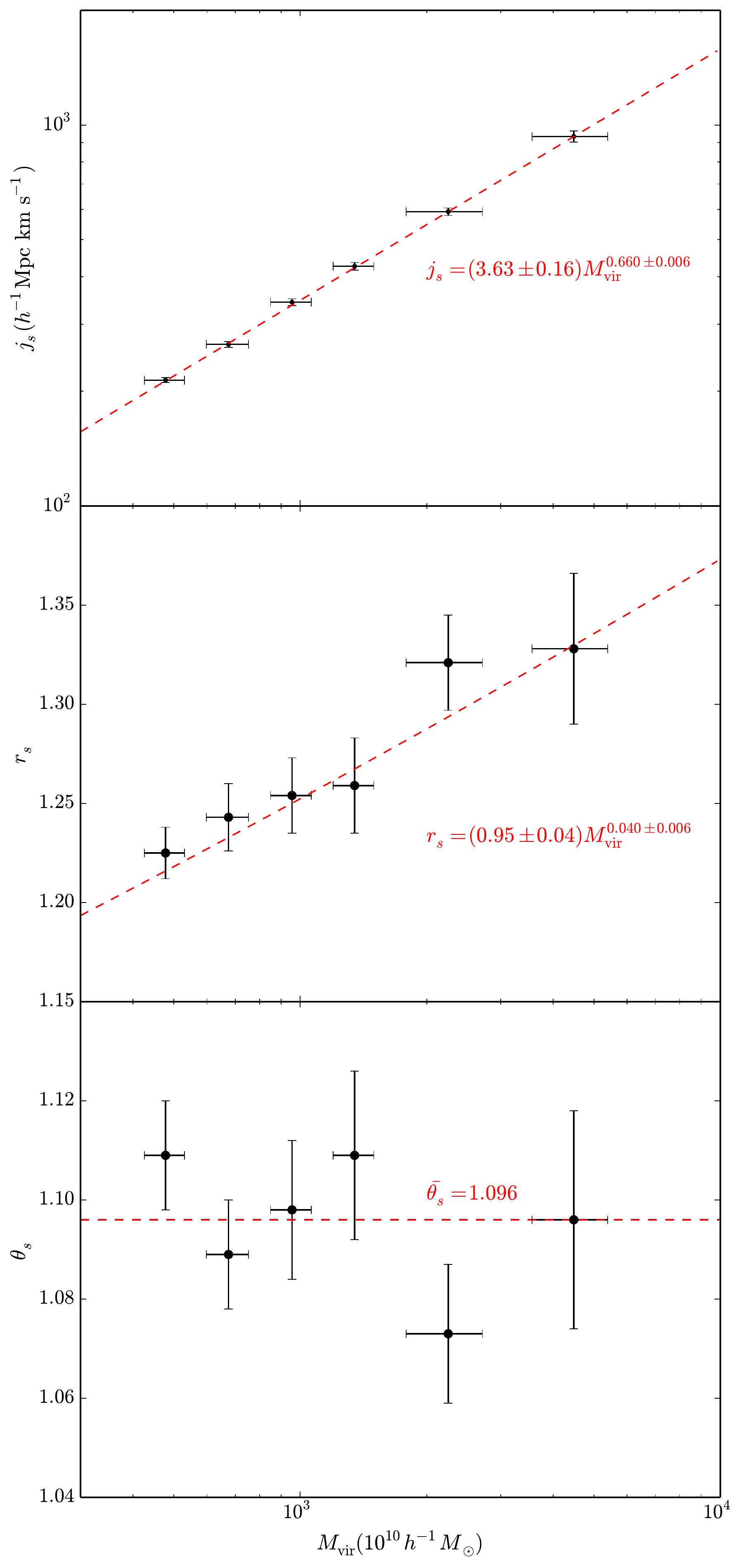}
\caption{Mass dependence of the fitted parameters $j_s, r_s,$ and $\theta_s$. The dashed line in the upper (middle) panel shows the best-fit $j_s-M_\mathrm{vir}$ $(r_s-M_\mathrm{vir})$ relation, while the dashed line in the lower panel presents the mean value $\bar{\theta}_s$ of six fitted $\theta_s$. The $x-$error bars show the standard deviation of halo mass in each mass bin, while the $y-$errors illustrate the jackknife estimated standard errors.}
\label{mass_dependence}
\end{figure}

The dependences on mass for $j_s, r_s,$ and $\theta_s$ are summarised in Figure \ref{mass_dependence}. The parameter $j_s$ scales with the halo mass as
\begin{equation}
j_s = (3.63\pm 0.16) M_\mathrm{vir}^{0.660\pm 0.006},
\end{equation}
which is consistent with the angular momentum - mass relation for virialized dark matter haloes, $j\propto M_\mathrm{vir}^{2/3}$ \citep[see][and references therein]{liao2015}. There is a weak trend for $r_s$ to increase for more massive haloes. The $r_s - M_\mathrm{vir}$ relation can be approximately described by $r_s \propto M_\mathrm{vir}^{0.040\pm 0.006}$. For $\theta_s$, there is no obvious trend for it to depend on halo masses. It fluctuates around a mean value of $1.096$, and agrees with this mean value in $\approx 1\sigma$ level.

Note that the mass dependences of the fitted parameters are not sensitive to choices of halo definition. We have performed a parallel analysis by redefining haloes with another common definition, i.e., $\Delta_\mathrm{vir}=200$, and the results are similar to those outlined above.

It is possible to eliminate the mass dependence of $j_s$ by introducing $j_s^\prime \equiv j_s/M_\mathrm{vir}^{2/3}$, and rewriting Equation (\ref{eq_fitted_amp}) as
\begin{equation}
j_\mathrm{mod}(r,\theta)=j_s^\prime M_\mathrm{vir}^{2/3}\frac{(r/r_s)^2}{(1+r/r_s)^4}\sin^2(\theta/\theta_s).
\end{equation}
In this case, all of the three parameters, $j_s^\prime$, $r_s$ and $\theta_s$, can be approximately regarded as universal for haloes with different masses. This is useful for roughly modelling the angular momentum distribution in a halo. But in the following discussions, we still use the model of Equation (\ref{eq_fitted_amp}).

The fitted profile tells us that the outer parts close to the equatorial plane of a halo tend to have larger specific angular momentum, while the inner parts near the polar direction usually have smaller $j$. Our results confirm the conclusion from 4 haloes in \citet[]{bullock2001}.

Similar to the scale radius in the NFW density profile, the parameters $r_s$ and $\theta_s$ in $j(r,\theta)$ measure the ``concentration'' of the specific angular momentum along $\pmb{\hat{r}}$ and $\pmb{\hat{\theta}}$ directions. The smaller $r_s$ $(\theta_s)$ is, the more $j$ concentrates to the halo centre ($z-$axis) along the radial ($-\pmb{\hat{\theta}}$) direction. The parameter $j_s$ measures the magnitude of $j$, which strongly correlates with the halo mass.

\begin{figure}
\includegraphics[width=250pt]{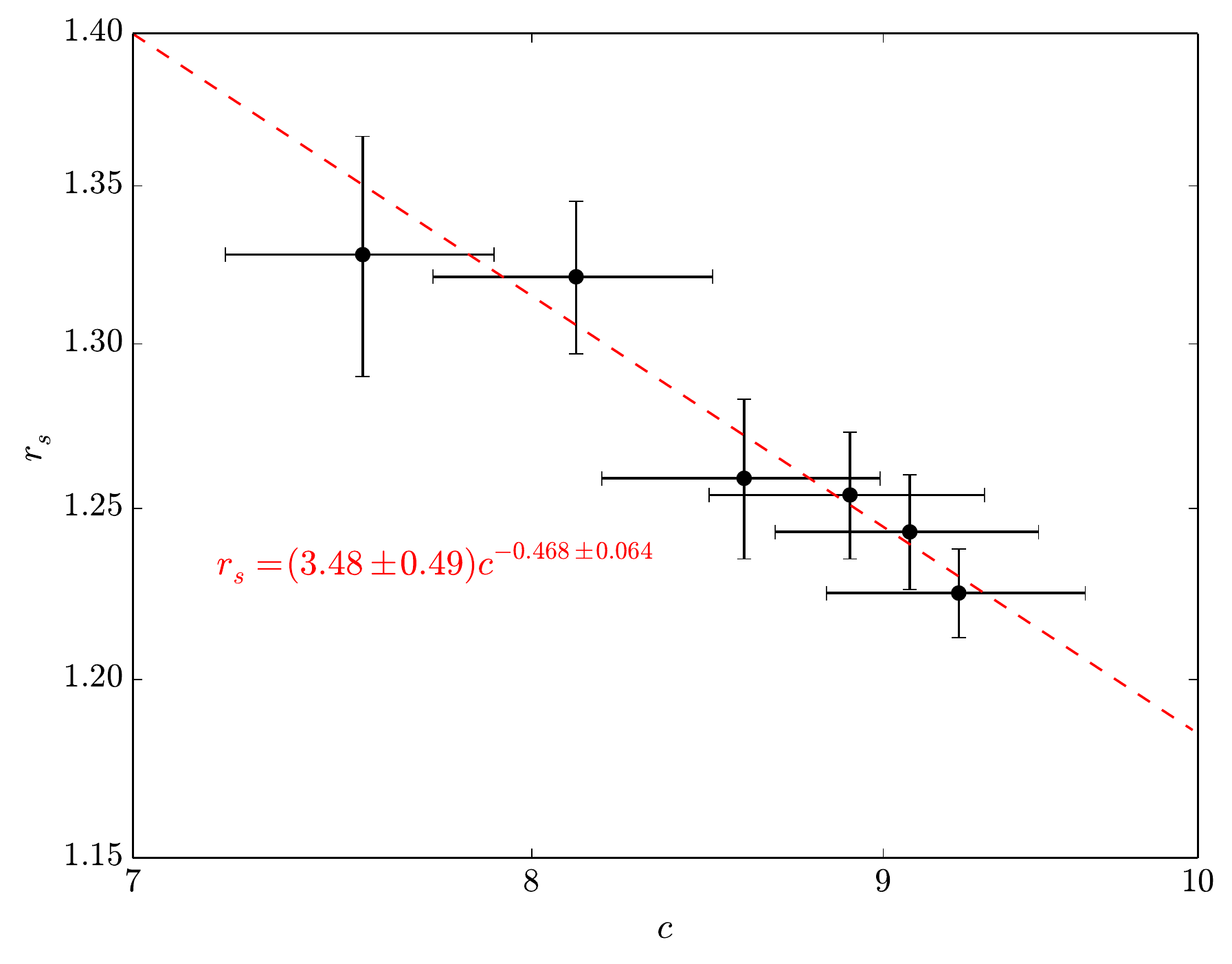}
\caption{The relation between the fitted parameter $r_s$ and halo concentration $c$. The halo concentrations $c$ are computed by fitting the stacked density profile with an NFW profile in each mass bin. The red dashed line marks the best-fit power law relation between $r_s$ and $c$, i.e., $r_s=(3.48 \pm 0.49)c^{-0.468 \pm 0.064}$.}
\label{rs_c_relation}
\end{figure}

The link between the fitted parameter $r_s$ in our profile and the scale radius $R_s$ in the NFW profile can be established by using the relation between $r_s$ and the halo concentration $c\equiv R_\mathrm{vir}/R_s$, which is a proxy of $R_s$. Since both $r_s$ and $c$ depend on halo mass as power laws, i.e., the $r_s-M_\mathrm{vir}$ relation and $c-M_\mathrm{vir}$ relation, there is also a power law relation between $r_s$ and $c$. For our halo sample, the $r_s-M_\mathrm{vir}$ relation is $r_s=(0.95 \pm 0.04)M^{0.040\pm 0.006}_\mathrm{vir}$, while the $c-M_\mathrm{vir}$ relation is $c=(16.60\pm 0.77)M^{-0.092\pm 0.007}_\mathrm{vir}$. Thus we expect that $r_s$ depends on $c$ as $r_s \approx 3.22c^{-0.435}$. This is confirmed by fitting directly the $r_s-c$ relation for our halo sample, i.e., $r_s=(3.48 \pm 0.49)c^{-0.468 \pm 0.064}$ (see Figure \ref{rs_c_relation}).

The universal specific angular momentum profile for simulated haloes is clearly different from that of a rigid body (RB). Rotating with a constant angular velocity $\omega_0$, a rigid body, which was usually adopted in early disk galaxy formation models \citep[e.g.,][]{mestel1963, dalcanton1997}, has a profile of $j_\mathrm{RB}(r,\theta)=\omega_0 R_\mathrm{vir}^2 r^2\sin^2\theta$. A visualized comparison can be found in panels (a) and (c) of Figure \ref{profile_contour}. However, as we outline in the following section, the universal $j(r,\theta)$ is quite similar to the one from the rigid shell (RS) model which is a modification of the RB model; see panel (d) of Figure \ref{profile_contour}.

\begin{table*}
 \centering
 \begin{minipage}{140mm}
  \caption{Best-fitted universal specific angular momentum profile for the Bolshoi haloes.}\label{bolshoi_amp}
  \begin{tabular} {@{}ccccccc@{}}
  \hline
    Mass bins & Halo number & Mean mass & $j_s$ & $r_s$ & $\theta_s$ & $\chi^2/\mathrm{d.o.f}$\\
    $(10^{12}h^{-1}M_\odot)$ & $N_h$ & $(10^{12}h^{-1}M_\odot)$ & $(h^{-1}\mathrm{Mpc}$ $\mathrm{km}$ $\mathrm{s}^{-1})$ & & & \\
   \hline
   $[4, 4\sqrt{2}]$ & 4218 & $4.78\pm 0.01$ & $214.0\pm 3.2 $ & $1.225\pm 0.013$ & $1.109\pm 0.011$ & 1.00 \\ 
   $[4\sqrt{2}, 8]$ & 3011 & $6.75\pm 0.02$ & $265.6\pm 4.6 $ & $1.243\pm 0.017$ & $1.089\pm 0.011$ & 0.90 \\ 
   $[8, 8\sqrt{2}]$ & 2170 & $9.56\pm 0.03$ & $342.6\pm 7.1$ & $1.254\pm 0.019$ & $1.098\pm 0.014$ & 1.07 \\ 
   $[8\sqrt{2}, 16]$ & 1518 & $13.47\pm 0.04$ & $425.7\pm 10.0$ & $1.259\pm 0.024$ & $1.109\pm 0.017$ & 1.23 \\ 
   $[16, 32]$ & 1849 & $22.50\pm 0.11$ & $592.4\pm 13.4$ & $1.321\pm 0.024$ & $1.073\pm 0.014$ & 1.24 \\ 
   $[32, 64]$ & 840 & $44.78\pm 0.32$ & $933.4\pm 31.6$ & $1.328\pm 0.038$ & $1.096\pm 0.022$ & 1.07 \\ 
\hline
\end{tabular}
\end{minipage}
\end{table*}

\section{Rigid shell model}\label{sec_model}

In the rigid shell model, we assume that a spherical shell with radius $R$ of a halo rotates rigidly along the $+z$-direction with velocity $v(R)=\sqrt{GM(<R)/R}$, where $M(<R)$ is the mass enclosed by the shell, and $G$ is the gravitational constant. The idea of rigid shells is similar to the onion-like mass growth picture for cold dark matter haloes \citep[][]{wang2011}. The specific angular momentum profile of a rigid shell halo is
\begin{equation}
j_\mathrm{RS}^\prime(r,\theta)=\sqrt{rR_\mathrm{vir}GM(<r)}\sin^2\theta.
\end{equation} 
The spin parameter $\lambda_\mathrm{RS}$ of this rigid shell halo, which rotates regularly and coherently, is much larger than that of a simulated halo which contains large amount of random motions. To compare with the $j(r,\theta)$ of simulated haloes, we rescale the profile of the rigid shell model to
\begin{equation}
j_\mathrm{RS}(r,\theta) = \frac{\lambda_\mathrm{sim}}{\lambda_\mathrm{RS}}j_\mathrm{RS}^\prime(r,\theta),
\end{equation} 
where the spin parameter of the rigid shell halo is
\begin{equation}
\lambda_\mathrm{RS}=\frac{R_\mathrm{vir}^2\int j_\mathrm{RS}^\prime(r,\theta)\rho(r)r^2\sin\theta drd\theta d\phi}{\sqrt{2}M_\mathrm{vir}V_\mathrm{vir}},
\end{equation}
and $\lambda_\mathrm{sim}$ is the spin parameter of the simulated halo. Here we adopt the definition of the spin parameter advocated by \citet[]{bullock2001}.

The specific angular momentum profile of a RS halo with an NFW density profile is
\begin{equation}\label{eq_amp_nfw}
j_\mathrm{RS,NFW}(r,\theta) = j_0 g(r) h(\theta),
\end{equation}
where the coefficient
\begin{equation}
j_0 = \frac{\lambda_\mathrm{sim}}{\lambda_\mathrm{RS,NFW}}\left(\frac{3}{4\pi\Delta_\mathrm{vir}\rho_\mathrm{cri}} \right)^{1/6} \sqrt{\frac{G}{f(c)}} M_\mathrm{vir}^{2/3},
\end{equation}
the radial part
\begin{equation}
g(r)= \sqrt{rf(cr)},
\end{equation}
the angular part
\begin{equation}
h(\theta)=\sin^2\theta,
\end{equation}
and $c$ is the concentration of the halo,
\begin{equation}
\lambda_\mathrm{RS,NFW} = \frac{\sqrt{2}}{3}\frac{c^3}{[f(c)]^{3/2}}\int_0^1 \sqrt{rf(cr)}\frac{r^2}{cr(1+cr)^2}dr,
\end{equation}
with
\begin{equation}
f(x) = \ln (1+x)-\frac{x}{1+x}.
\end{equation}
The results for the Einasto density profile \citep[]{merritt2006} can be found in Appendix \ref{appendix_einasto}.

\begin{figure}
\includegraphics[width=262pt]{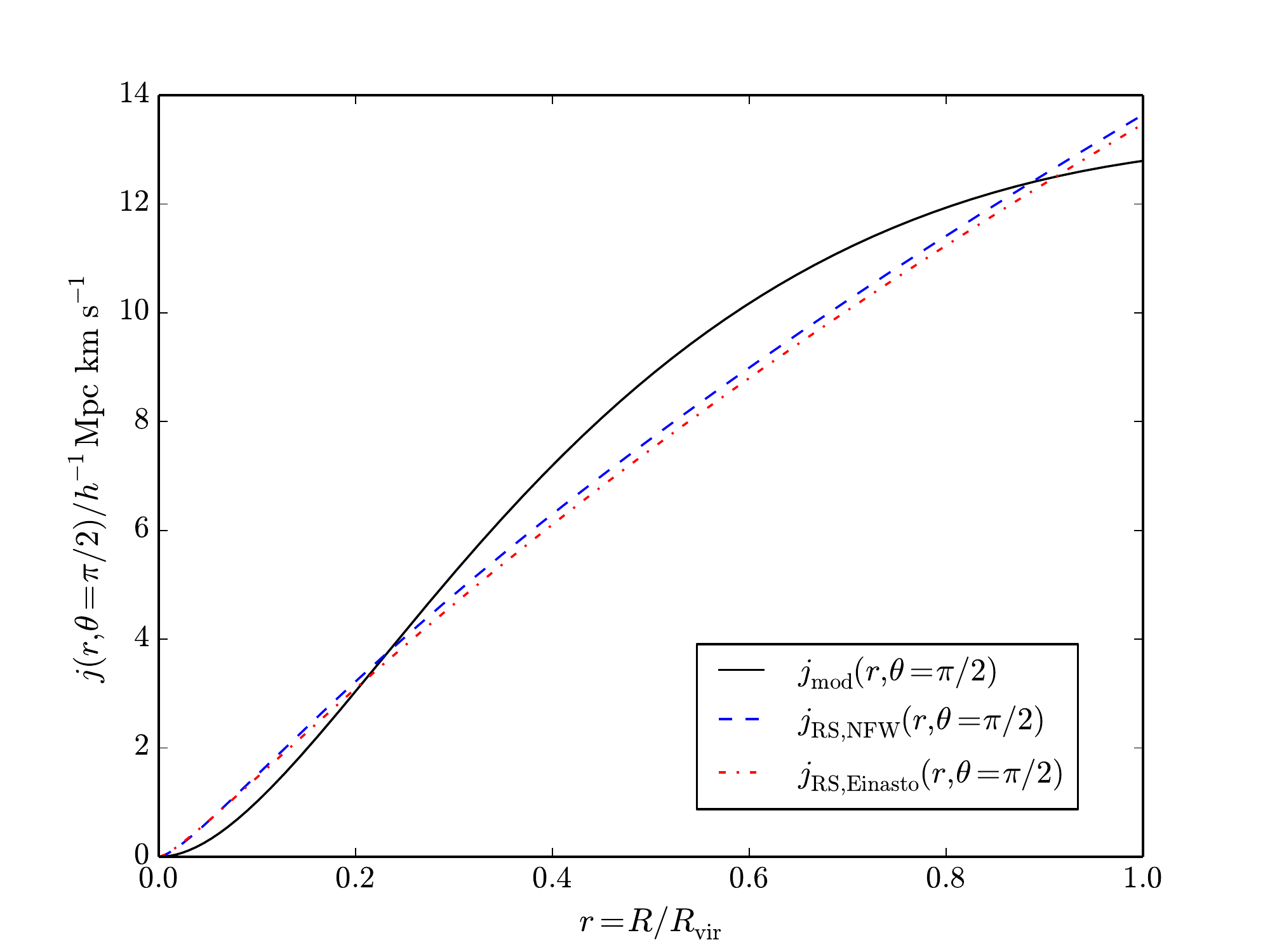}
\caption{Comparisons between the specific angular momentum profiles from the fitting and rigid shell model. The fitted $j_\mathrm{mod}(r,\theta=\pi/2)$ for haloes with $M_\mathrm{vir}=[4,4\sqrt{2}]\times 10^{12}$ $h^{-1}M_\odot$ is shown with a solid line. $j(r,\theta=\pi/2)$ calculated from the rigid shell model with the NFW and Einasto density profiles are plotted as the dashed and dash-dotted lines respectively. The input parameters, $\lambda_\mathrm{sim}=0.035, c=9.23, A=17.05,$ and $\alpha=0.170$, are computed from the simulated halo sample.}
\label{rigid_shell}
\end{figure}

From Equation (\ref{eq_amp_nfw}), we can find that $j_\mathrm{RS}(r,\theta)$ predicts similar behaviours as the fitted $j_\mathrm{mod}(r,\theta)$: (i) the coefficient $j_0$ is proportional to $M_\mathrm{vir}^{2/3}$; (ii) the angular part $h(\theta)$ is the square of a sine function; (iii) the radial part $g(r)$ increases as $r$ in a qualitatively similar way, as shown in Figure \ref{rigid_shell}. This leads to the similarity between panels (a) and (d) in Figure \ref{profile_contour}.  

Of course, the rigid shell model should not be an exact model for simulated haloes since its rotation is too regular. But it provides us an effective physical picture to understand qualitatively the fitted profile from simulated haloes. The exact origin of the universal profile remains to be understood.

\section{Links to baryonic processes}\label{sec_obs}
In classical theories of disk galaxy formation, it is assumed that the gas shares the same specific angular momentum distribution as dark matter, and conserves its angular momentum in the cooling process to form a disc finally~\citep[][]{mestel1963, fall1980, mo1998}. However, this picture leads to an excess of low angular momentum material and thus too compact a density profile for galactic discs \citep[][]{bullock2001, bosch2001a, bosch2001b}. One of the possible resolutions is that the galactic winds from supernova (SN) feedbacks remove gas with low angular momentum \citep[][]{binney2001}. Our profile implies that such low angular momentum material locates in the galactic center and a conical region along the polar direction \citep[see][for a similar conclusion based on individual merger simulations]{sharma2012}, and thus a feasible mechanism should prohibit the gas in these regions from forming the disc. Recent numerical simulations have confirmed that SN feedbacks preferentially blow away gas with low angular momentum \citep[see e.g.][]{gavernato2010, brook2011, guedes2011}. Especially, \citet[][]{brook2011} clearly show that the path of the outflows is along the polar direction.

\begin{figure}
\includegraphics[width=245pt]{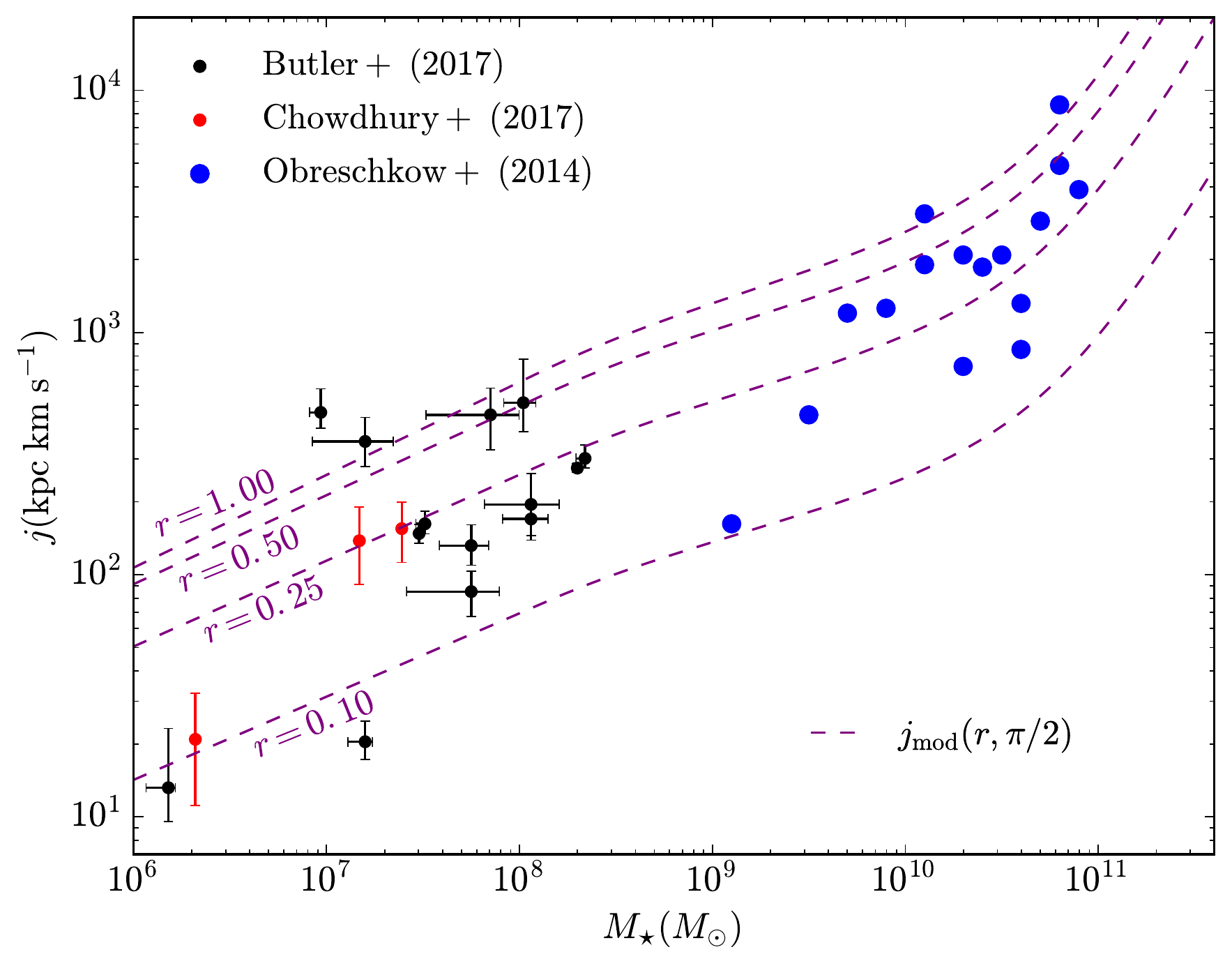}
\caption{Comparison of the $j-M_{\star}$ relation for dark matter and HI gas. For each theoretical curve (dashed line), we fix the numerical values of $r$ and $\theta$ and then calculate $j_{\mathrm{mod}}$ for different stellar mass $M_{\star}$. From top to bottom, the values for $r$ are 1.0, 0.5, 0.25, and 0.1 accordingly while $\theta$ is always $\pi/2$. Appendix \ref{details_figure_7} details the evaluation procedure. Observation data for the average specific angular momentum of HI gas $j_{\mathrm{gas}}$ and stellar mass $M_{\star}$ are taken from the following literatures: 16 blue points from \citet[][]{obreschkow2014}, 14 black points from \citet[][]{butler2017}, and 3 red points from \citet[][]{chowdhury2017}.}
\label{Observation}
\end{figure}

Our profile suggests that the specific angular momentum at any spatial position within a halo scales approximately with the halo mass as $j_\mathrm{mod}(r,\theta)\propto M^{2/3}_\mathrm{vir}$. Before gas cooling, the average specific angular momentum of gas locating at volume $V$, $j_\mathrm{gas} \approx \int_V j_\mathrm{mod}(r,\theta) \rho_\mathrm{gas}(r)dV / \int_V \rho_\mathrm{gas}(r)dV$, is expected to follow a similar mass scaling, $j_\mathrm{gas} \propto M^{2/3}_\mathrm{vir}$, assuming gas shares a similar specific angular momentum profile as dark matter. It is interesting to see how this mass scaling relation changes after gas cooling and condensation. In Figure \ref{Observation}, we compare the observed $j_\mathrm{gas} - M_\star$ relations of HI gas \citep[][]{obreschkow2014, butler2017, chowdhury2017} and the $j_\mathrm{mod}(r,\theta)-M_\star$ scaling at different positions. Notice that in order to have a direct comparison with observations, here we adopt the stellar mass of a galaxy, $M_\star$, and use the stellar mass - halo mass relation \citep[e.g.][]{behroozi2013, moster2013, kravtsov2014} to convert $M_\mathrm{vir}$ into $M_\star$; see Appendix \ref{details_figure_7} for details. We find that the observed $j_\mathrm{gas}-M_\star$ relation has the same universal shape as $j_\mathrm{mod}(r,\theta)-M_\star$ relation (at any $r$ and $\theta$). This implies that although the detailed processes are fairly complicated (e.g., cooling, feedbacks, torques, etc.), the HI gas condenses out from different parts of a halo in a way that is scale-free, i.e., it does not depend on halo masses. The similarity between the observed $j_\mathrm{gas} - M_{\star}$ relation and our $j_\mathrm{mod}(r,\theta)-M_\star$ curves also supports the explanation of a higher intercept of $j_{\mathrm{gas}} - M_{\star}$ relation compared to the stellar component $(j_{\star} - M_{\star})$ within the CDM framework~\citep[][]{butler2017}. The stars mainly form from the gas in more central parts which have lower angular momenta and thus occupy a lower position in the $j-M_{\star}$ plane.

The specific angular momentum distribution of a dark matter halo according to our universal model, $j(r,\theta)$, can be easily calculated once $M_{\text{vir}}$ is known. Using the stellar mass - halo mass relation, one can infer $M_{\text{vir}}$ observationally from a galaxy's stellar mass $M_{\star}$. Therefore the specific angular momentum distribution for a dark matter halo, on average, can be obtained by weighing its stars. Analytical \citep[e.g.][]{blumenthal1986} as well as numerical studies \citep[e.g.][]{dutton2016} show that galaxy formation processes can alter the inner dark matter density profile predicted by dark-matter-only N-body simulations. Then $j(r,\theta)$ may also be distorted near a galaxy's center. However, towards the outskirts of galaxies, star formation and feedbacks are less important, and our predicted angular momentum profile should be comparable to observation.

\section{Discussions}\label{sec_dis}

In this article, we use high resolution haloes in the Bolshoi simulation to show the existence of a universal stacked spatial profile for the angular momenta of haloes, $j(r,\theta)=j_s \sin^2(\theta/\theta_s) (r/r_s)^2/(1+r/r_s)^4$ with three parameters, $j_s, r_s$ and $\theta_s$. We show that $j_s$ strongly correlates with halo mass $M_\mathrm{vir}$ as a power law, $j_s\propto M_\mathrm{vir}^{2/3}$, which is related to the angular momentum - mass relation. The parameter $r_s$ depends weakly on $M_\mathrm{vir}$, and $\theta_s$ is independent of the halo mass. This axisymmetric profile is an improvement of previous spherically symmetric angular momentum profiles, since the angular momentum itself defines a special direction. Although the origin of this newly found profile remains unknown, we show that it is similar to the one from the rigid shell model.

The angular momentum profile, $j(r,\theta)$, encrypts both the position and velocity information of particles in a virialized dark matter halo. Like the universal density profile, it represents another equilibrium property of a collisionless N-body system. Understanding its origin will offer us more insights on the structure formation theory.

The fitting profile of $j(r,\theta)$ is useful in modelling haloes' angular momenta. With $j(r,\theta)$ and the density profile $\rho(r)$, we can calculate many related quantities directly, such as the total angular momentum $J$, spin parameter $\lambda$, angular momentum - mass relation, spherical angular momentum profiles $j_d(r)$ and $j_c(<r)$, etc. 

\begin{figure}
\includegraphics[width=245pt]{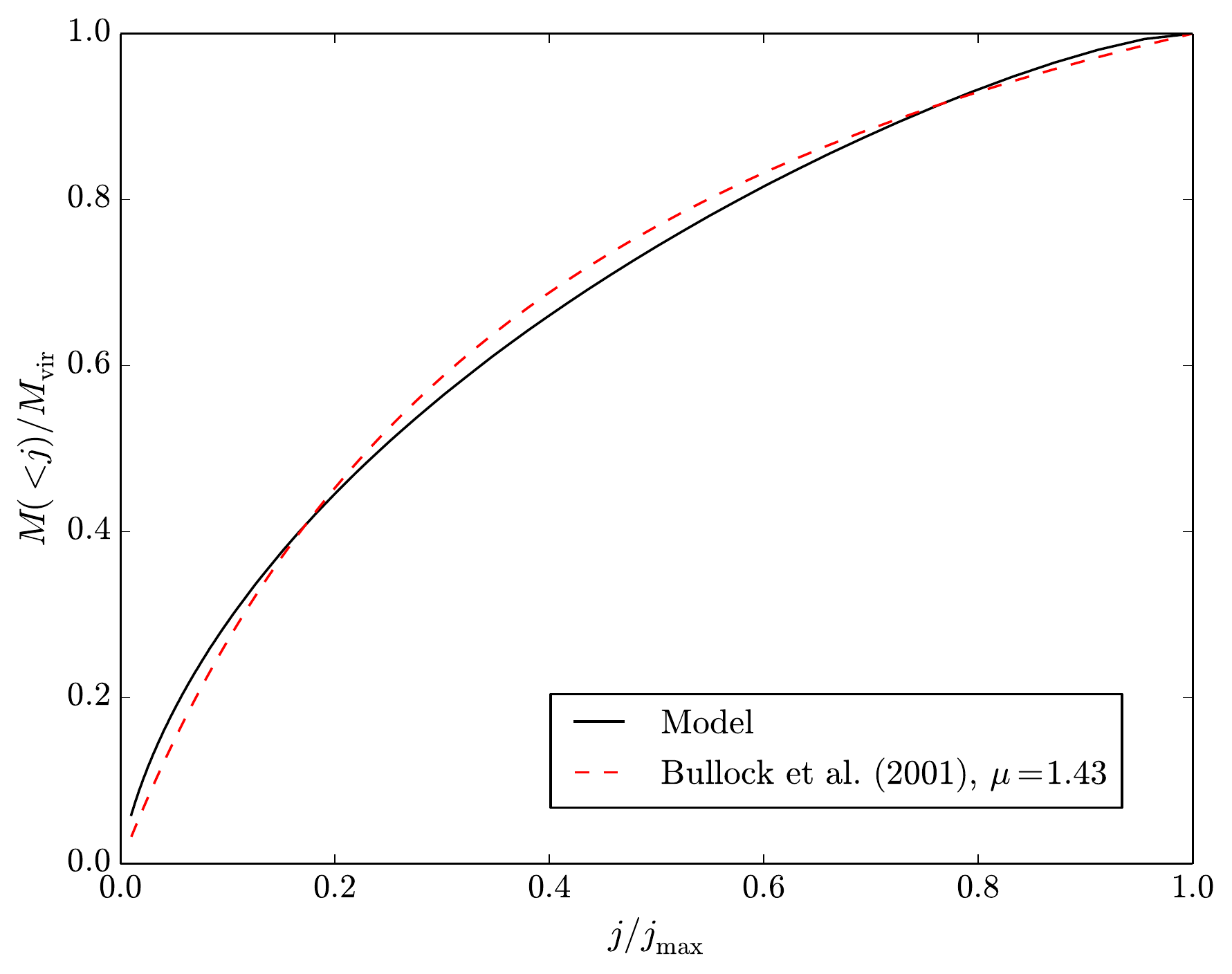}
\caption{Mass profiles of specific angular momenta from our fitted $j_\mathrm{mod}(r,\theta)$ with the NFW density profile (solid) using the parameters of the stacked haloes with $M_\mathrm{vir}=[4,4\sqrt{2}]\times 10^{12}$ $h^{-1}M_\odot$, and the empirical function proposed by \citet[]{bullock2001} with $\mu=1.43$ (dashed).}
\label{mass_profile}
\end{figure}

As an example, we use $j_\mathrm{mod}(r,\theta)$ to discuss the mass profile of angular momentum, $M(<j)$. As shown in Figure \ref{mass_profile}, $M(<j)$ calculated from our fitted $j_\mathrm{mod}(r,\theta)$ with an NFW density profile agrees qualitatively well with the empirical fitting function given by \citet[]{bullock2001},
\begin{equation}
M_{\rm{emp}}(<j)=M_\mathrm{vir}\frac{\mu j/j_\mathrm{max}}{\mu-1 + j/j_\mathrm{max}},
\end{equation}
with $\mu=1.43$ which is close to the mean value of $1.25$. Although both $j(r,\theta)$ and $M_{\rm{emp}}(<j)$ are empirically fitted from numerical simulations, we cannot expect $M(<j)$ calculated from $j(r,\theta)$ to be exactly the same as $M_{\mathrm{emp}}(<j)$ since $j(r,\theta)$ is obtained by stacking haloes with similar masses while $M_{\rm{emp}}(<j)$ is computed from individual haloes which have larger noises.

As shown in Section \ref{sec_obs}, the $j-M_{\star}$ relation of dark matter haloes predicted by our profile and that obtained from the observational data on HI gas in disk galaxies share similar shapes. We also discuss that, once the stellar mass of a galaxy is given, we can compute the associated halo mass through the stellar mass - halo mass relation, and further specify the corresponding stacked angular momentum profile. Therefore we can infer the averaged spatial distribution of a dark matter halo's angular momentum, which cannot be observed directly, from the observation of its associated baryons. However, baryon physics should be included in more detailed modeling. The most recent cosmological hydrodynamic simulations start to reproduce galaxy morphology more realistically and can be used to study the dependence of galaxies' angular momenta on morphology \citep[e.g.][]{fiacconi2015, teklu2015, genel2015, zavala2016}. Investigations of the interplay between dark matter and baryons with future large cosmological hydrodynamic simulations will improve our understanding of the dark matter spatial angular momentum profile and its relation to the baryonic counterpart.

Here, we only study the specific angular momentum profile for haloes in the $\Lambda$CDM model through a dark matter-only simulation at $z=0$. It is worthwhile to study how this profile depends on the dark matter model (e.g. warm dark matter, self-interacting dark matter, etc.) and cosmological models at different redshifts. These will offer us more insights on the origin of the universal angular momentum profile discussed in this article.

\section*{Acknowledgements}
The authors thank Kristin Riebe for helps in querying data from the CosmoSim database. We are grateful for the comments and suggestions from the anonymous referee. We also acknowledge the inspiring discussion with Hsiang Hsu Wang. JC is thankful to Dalong Cheng and Yipeng Jing for their hospitality extended to him during a visit to the Center for Astronomy and Astrophysics (CAA) at Shanghai Jiao Tong University, where part of this work is done. This work is supported partially by an RGC Grant CUHK14301214. The CosmoSim database used in this paper is a service by the Leibniz-Institute for Astrophysics Potsdam (AIP). The Bolshoi simulations used in this paper have been performed within the Bolshoi project of the University of California High-Performance AstroComputing Center (UC-HiPACC) and were run at the NASA Ames Research Center. This research has made use of NASA's Astrophysics Data System.

\appendix

\section{Removing unbound particles}\label{appendix_unbound}
To remove unbound particles, we assume that the density of a halo distributes spherically and calculate the potential at the position of every particle,
\begin{equation}
\phi(R)=G\int_{0}^{R} \frac{M(<R^\prime)}{R^{\prime 2}}dR^\prime + \phi_0,
\end{equation}
where $R$ is radial position from the halo centre, and the constant
\begin{equation}
\phi_0 = -G\left[\frac{M_\mathrm{vir}}{R_\mathrm{vir}} + \int_0^{R_\mathrm{vir}}\frac{M(<R^\prime)}{R^{\prime 2}}dR^\prime \right].
\end{equation}
The escape velocity $v_\mathrm{e}(R)$ is defined as
\begin{equation}
v_\mathrm{e}(R)=\sqrt{2|\phi(R)|}.
\end{equation}
If a particle has velocity $v_i > v_\mathrm{e}$, then it is unbound and is removed from the particle list of the halo. After removing all unbound particles, the position of the centre of mass of the new particle set is calculated. 

We iteratively remove unbound particles in the new particle set and calculate the new centre-of-mass position, until $\leq 3$ particles are removed or the iteration number $N_\mathrm{iter}\geq 10$. The iteration numbers for our halo sample approximately follow a lognormal distribution with a median of $\sim 4$, and variance of $\sim 0.4$. 

Our halo sample after removing unbound particles agrees well with the Bolshoi halo sample which uses a different removing scheme \citep[]{klypin1997}. For example, $93\%$ of our haloes have their masses differing from the Bolshoi counterparts by smaller than $5\%$, i.e. $|M_\mathrm{vir}-M_\mathrm{vir,Bolshoi}|/M_\mathrm{vir,Bolshoi}<5\%$.

\section{Discreteness effects and spatial binning schemes}\label{appendix_fitting}
To see the effects of the spatial binning scheme and particle resolution $N_\mathrm{th}$ on the angular momentum profile, we divide the mass bin of $[32,64]\times 10^{12}h^{-1}M_\odot$ using different spatial binning schemes and different resolutions $N_\mathrm{th}$, and we compare the fitted $j(r,\theta)$. The results are presented in Table \ref{fitting_tests}. We conclude that in a large range of $N_\mathrm{th}$ and different spatial binning schemes, the fitted $j-$profile is not affected.

\begin{table*}
 \centering
 \begin{minipage}{140mm}
  \caption{Best-fitted angular momentum profile for haloes in the mass bin of $[32,64]\times 10^{12}$ $h^{-1}M_\odot$.}\label{fitting_tests}
  \begin{tabular} {@{}ccccccc@{}}
  \hline
    $N_\mathrm{th}$ & Number of $r$ bins & Number of $\theta$ bins & $j_s$ & $r_s$ & $\theta_s$ & $\chi^2/\mathrm{d.o.f}$\\
     & &  & $(h^{-1}\mathrm{Mpc}$ $\mathrm{km}$ $\mathrm{s}^{-1})$ & & & \\
   \hline
   100 & 49 & 48 & $927.4\pm 31.1$ & $1.320\pm 0.037$ & $1.097\pm 0.023$ & 0.99 \\
   200 & 34 & 34 & $933.4\pm 31.6$ & $1.328\pm 0.038$ & $1.096\pm 0.022$ & 1.07 \\ 
   400 & 24 & 24 & $933.9\pm 31.8$ & $1.327\pm 0.039$ & $1.096\pm 0.022$ & 1.12 \\ 
   800 & 17 & 17 & $928.6\pm 31.5$ & $1.320\pm 0.039$ & $1.094\pm 0.022$ & 1.23 \\ 
   1600 & 12 & 12 & $922.6\pm 31.5$ & $1.310\pm 0.039$ & $1.092\pm 0.023$ & 1.40 \\ 
\hline
\end{tabular}
\end{minipage}
\end{table*}

\section{Rigid shell model with Einasto density profile}\label{appendix_einasto}
The Einasto density profile \citep[]{merritt2006}, $\rho(r)=\rho_0\exp(-Ar^\alpha)$, is shown to fit the stacked halo density profile better than the NFW one \citep[]{gao2008}. For a rigid shell halo with an Einasto density profile, the specific angular momentum profile can be calculated to be
\begin{eqnarray}\label{eq_amp_einasto}
j_\mathrm{RS,Einsto}(r,\theta)&=&\frac{\lambda_\mathrm{sim}}{\lambda_\mathrm{RS,NFW}}\left(\frac{3}{4\pi\Delta_\mathrm{vir}\rho_\mathrm{cri}} \right)^{1/6} \times  \\ \nonumber
&&\sqrt{\frac{G}{\gamma(3/\alpha,A)}}M_\mathrm{vir}^{2/3}\sqrt{r\gamma(3/\alpha,Ar^\alpha)}\sin^2\theta,
\end{eqnarray}
where $\gamma(s,x)=\int_0^x t^{s-1}e^{-t}dt$ is the lower incomplete gamma function, and
\begin{eqnarray}
\lambda_\mathrm{RS,Einasto} &=& \frac{\sqrt{2}}{3}\frac{\alpha A^{3/\alpha}}{[\gamma(3/\alpha, A)]^{3/2}}\times \\ \nonumber
&&\int_0^1 \sqrt{r\gamma(3/\alpha, A)}\exp(-Ar^\alpha)r^2 dr.
\end{eqnarray}
An illustration of $j_\mathrm{RS,Einsto}(r,\theta)$ can be found in Figure \ref{rigid_shell}.

\section{Comparing specific angular momentum - mass relations between dark matter and gas}\label{details_figure_7}
To obtain the $j_\mathrm{mod}(r,\theta)-M_\star$ relation, we need to convert the halo mass, $M_\mathrm{vir}$, into a corresponding stellar mass, $M_\star$. This can be done with the stellar mass - halo mass relation,
\begin{equation}\label{eq:kra14a}
           \log_{10}\left(M_{\star}\right) = \log_{10}(\epsilon M_{1}) + f\left(\log_{10}\left(\frac{M_{\mathrm{vir}}}{M_{1}}\right)\right) - f(0),          
\end{equation}
with
\begin{equation}\label{eq:kra14b}
             f(x) = -\log_{10}\left(10^{\alpha x} + 1\right) + \delta\frac{\left[\log_{10}\left(1 + \exp{(x)}\right)\right]^{\gamma}}{1 + \exp{\left(10^{-x}\right)}},
\end{equation}
where masses are measured in units of solar mass $M_{\odot}$. There are five parameters: $M_{1}$, $\epsilon$, $\alpha$, $\delta$, and $\gamma$. Their values used in our calculation are taken from \citet[][]{kravtsov2014} and are listed in Table \ref{table_params}. For a given $M_{\star}$, we use Equation (\ref{eq:kra14a}) to derive its corresponding $M_{\mathrm{vir}}$ and then apply our model to calculate the dark matter halo's specific angular momentum profile:
\begin{equation}
             j_{\mathrm{mod}}(r,\theta) = j_s\frac{\left(r/r_s\right)^2}{\left(1 + r/r_s\right)^4}\sin^2{\left(\theta/\theta_s\right)}, 
\end{equation}
with
\begin{equation}
\frac{j_{s}}{h^{-1}\phantom{0}\text{Mpc}\phantom{0}\text{km}\phantom{0}\text{s}^{-1}} = A_{\text{vir}}\left(\frac{M_{\text{vir}}}{10^{10}h^{-1}M_{\odot}}\right)^{B_{\text{vir}}},
\end{equation}
\begin{equation}
r_{s} = E_{\text{vir}}\left(\frac{M_{\text{vir}}}{10^{10}h^{-1}M_{\odot}}\right)^{F_{\text{vir}}},
\end{equation}
and
\begin{equation}
\theta_s = 1.096.
\end{equation}
Values for parameters $A_{\mathrm{vir}}$, $B_{\mathrm{vir}}$, $E_{\mathrm{vir}}$, and $F_{\mathrm{vir}}$ are also summarized in Table \ref{table_params}.

\begin{table*}
\renewcommand\arraystretch{1.5} 
 \centering
  \caption{Model parameters for the specific angular momentum profile}\label{table_params}
  \begin{tabular} {@{}ccccccccc@{}}
  \hline
    $A_{\mathrm{vir}}$ & $B_{\mathrm{vir}}$ & $E_{\mathrm{vir}}$& $F_{\mathrm{vir}}$ & $\log_{10}(M_{1}/M_{\star})$ & $\log_{10}{\epsilon}$ & $\alpha$ & $\delta$ & $\gamma$ \\
   \hline
    3.63 & 0.660 & 0.95 & 0.040 & 11.39 & -1.685 & -1.740 & 4.335 & 0.531\\
\hline
\end{tabular}
\end{table*}


\begin{thebibliography}{101}
\bibitem[Barns \& Efstathiou(1987)]{barns1987} Barns, J., \& Efstathiou, G. 1987, ApJ, 319, 575
\bibitem[Behroozi et al.(2013)]{behroozi2013} Behroozi, P.~S., Wechsler, R.~H., \& Conroy, C.\ 2013, \apj, 770, 57 
\bibitem[Bett et al.(2010)]{bett2010} Bett, P., Eke, V., Frenk, C. S., Jenkins, A., \& Okamoto, T. 2010, MNRAS, 404, 1137
\bibitem[Binney et al.(2001)]{binney2001} Binney, J., Gerhard, O., \& Silk, J. 2001, MNRAS, 321, 471
\bibitem[Blumenthal et al.(1986)]{blumenthal1986} Blumenthal, G.~R., Faber, S. M., Flores, R., \& Primack, J. R. 1986, ApJ, 301, 27
\bibitem[Brook et al.(2011)]{brook2011} Brook, C. B., Governato, F., Ro\u{s}kar, R., et al. 2011, MNRAS, 415, 1051
\bibitem[Bryan \& Norman(1998)]{bryan1998} Bryan, G. L., \& Norman, M. L. 1998, ApJ, 495, 80
\bibitem[Bullock et al.(2001)]{bullock2001} Bullock, J. S., Dekel, A., Kolatt, T. S., et al. 2001, ApJ, 555, 240
\bibitem[Butler et al.(2017)]{butler2017} Butler, K.~M., Obreschkow, D., \& Oh, S.-H.\ 2017, \apjl, 834, L4 
\bibitem[Chen \& Jing(2002)]{chen2002} Chen, D. N., \& Jing, Y. P. 2002, MNRAS, 336, 55
\bibitem[Chen et al.(2003)]{chen2003} Chen, D. N., Jing, Y. P., \& Yoshikaw, K. 2003, ApJ, 597, 35
\bibitem[Chowdhury \& Chengalur(2017)]{chowdhury2017} Chowdhury, A., \& Chengalur, J.~N.\ 2017, \mnras, 467, 3856 
\bibitem[Dalcanton et al.(1997)]{dalcanton1997} Dalcanton, J. J., Spergel, D. N., \& Summers, F. J. 1997, ApJ, 482, 659
\bibitem[Dutton et al.(2016)]{dutton2016} Dutton, A. A., Macci{\`o}, A. V., Dekel, A., et al. 2016, MNRAS, 461, 2658
\bibitem[Fall \& Efstathiou(1980)]{fall1980} Fall, S. M., \& Efstathiou, G. 1980, MNRAS, 193, 189
\bibitem[Fiacconi et al.(2015)]{fiacconi2015} Fiacconi, D., Feldmann, R., \& Mayer, L. 2015, MNRAS, 446, 1957
\bibitem[Gao et al.(2008)]{gao2008} Gao, L., Navarro, J. F., Cole, S., et al. 2008, MNRAS, 387, 536
\bibitem[Genel et al.(2015)]{genel2015} Genel, S., Fall, S. M., Hernquist, L., et al. 2015, ApJ, 804, L40
\bibitem[Governato et al.(2010)]{gavernato2010} Governato, F., Brook, C., Mayer, L., et al. 2010, Natur, 463, 203
\bibitem[Guedes et al.(2011)]{guedes2011} Guedes, J., Callegari, S., Madau, P., \& Mayer, L. 2011, ApJ, 742, 76
\bibitem[Hamaus et al.(2014)]{hamaus2014} Hamaus, N., Sutter, P. M., \& Wandelt, B. D. 2014, PhRvL, 112, 251302
\bibitem[Hayashi \& White(2008)]{hayashi2008} Hayashi, E., \& White, S. D. M. 2008, MNRAS, 388, 2 
\bibitem[Klypin \& Holtzman(1997)]{klypin1997} Klypin, A., \& Holtzman, J. 1997, arXiv:astro-ph/9712217
\bibitem[Klypin et al.(2011)]{klypin2011} Klypin, A. A., Trujillo-Gomez, S., \& Primack, J. 2011, ApJ, 740, 102
\bibitem[Kravtsov et al.(2014)]{kravtsov2014} Kravtsov, A., Vikhlinin, A., \& Meshscheryakov, A.\ 2014, arXiv:1401.7329
\bibitem[Levenberg(1944)]{levenberg1944} Levenberg, K. 1944, QApMa, 2, 164
\bibitem[Liao et al.(2015)]{liao2015} Liao, S., Cheng, D., Chu, M.-C., \& Tang, J. 2015, ApJ, 809, 64
\bibitem[Marquardt(1963)]{marquardt1963} Marquardt, D. 1963, Journal of the Society for Industrial and Applied Mathematics, 11, 431
\bibitem[Merritt et al.(2006)]{merritt2006} Merritt, D., Graham, A. W., Moore, B., Diemand, J., \& Terzi\'{c}, B. 2006, ApJ, 132, 2685
\bibitem[Mestel(1963)]{mestel1963} Mestel, L. 1963, MNRAS, 126, 553
\bibitem[Mo et al.(1998)]{mo1998} Mo, H. J., Mao, S., \& White, S.~D.~M.\ 1998, \mnras, 295, 319
\bibitem[Mo et al.(2010)]{mo2010} Mo, H., van den Bosch, F., \& White, S. 2010, Galaxy formation and evolution (New York: Cambridge University Press)
\bibitem[Moster et al.(2013)]{moster2013} Moster, B.~P., Naab, T., \& White, S.~D.~M.\ 2013, \mnras, 428, 3121 
\bibitem[Navarro et al.(1995)]{navarro1995} Navarro, J. F., Frenk, C. S., \& White, S. D. M. 1995, MNRAS, 275, 720
\bibitem[Navarro et al.(1996)]{navarro1996} Navarro, J. F., Frenk, C. S., \& White, S. D. M. 1996, ApJ, 462, 563
\bibitem[Navarro et al.(1997)]{navarro1997} Navarro, J. F., Frenk, C. S., \& White, S. D. M. 1997, ApJ, 490, 493
\bibitem[Obreschkow \& Glazebrook(2014)]{obreschkow2014} Obreschkow, D., \& Glazebrook, K.\ 2014, \apj, 784, 26
\bibitem[Reed et al.(2011)]{reed2011} Reed, D. S., Koushiappas, S. M., \& Gao, L. 2011, MNRAS, 415, 3177
\bibitem[Sch\"{a}fer(2009)]{schafer2009} Sch\"{a}fer, B. M. 2009, IJMPD, 18, 173
\bibitem[Sharma \& Steinmetz(2005)]{sharma2005} Sharma, S., \& Steinmetz, M. 2005, ApJ, 628, 21
\bibitem[Sharma et al.(2012)]{sharma2012} Sharma, S., Steinmetz, M., \& Bland-Hawthorn, J. 2012, ApJ, 750, 107
\bibitem[Teklu et al.(2015)]{teklu2015} Teklu, A. F., Remus, R.-S., Dolag, K., et al. 2015, ApJ, 812, 29
\bibitem[van den Bosch(2001)]{bosch2001a} van den Bosch, F. C. 2001, MNRAS, 327, 1334
\bibitem[van den Bosch et al.(2002)]{bosch2002} van den Bosch, F. C., Abel, T., Croft, R. A. C., Hernquist, L., \& White, S. D. M. 2002, ApJ, 576, 21
\bibitem[van den Bosch et al.(2001)]{bosch2001b} van den Bosch, F. C., Burkert, A., \& Swaters, R. A. 2001, MNRAS, 326, 1205
\bibitem[Wang et al.(2011)]{wang2011} Wang, J., Navarro, J. F., Frenk, C. S., et al. 2011, MNRAS, 413, 1373
\bibitem[Zavala et al.(2016)]{zavala2016} Zavala, J., Frenk, C. S., Bower, R., et al. 2016, MNRAS, 460, 4466

\end{thebibliography}
\end{document}